\documentclass[runningheads]{llncs}
\usepackage{graphicx} 
\usepackage{amsmath}                 
\usepackage{amssymb}  
\usepackage[accsupp]{axessibility}  
\usepackage[numbers]{natbib}
\usepackage{multirow}
\usepackage{mathabx}
\usepackage{hyperref}
\usepackage{geometry}
\usepackage{epigraph}
\usepackage{csquotes}
\usepackage{wrapfig}
\usepackage{xcolor}
\title{Re-assessing the evidence for mental rotation abilities in children using computational models}

\author{Arthur Aubret and Jochen Triesch}
\institute{Frankfurt Institute for Advanced Studies, Frankfurt-am-Main}
\newcommand{\review}[1]{\textcolor{black}{#1}}

\date{}

\begin{document}
\maketitle

\review{}

\section{Highlights}
\begin{itemize}
    \item We use a bio-inspired machine learning models to examine evidence of mental rotation abilities in children
    \item We \review{show} that a simple recognition strategy \review{suffices} to solve habituation-based tasks used to assess mental rotations in young children.
    \item We demonstrate that \review{a model} forming expectations about a rotation transformation can explain children's results in mental rotation tasks with the violation-of-expectation paradigm.
    \item The investigated recognition strategies \review{no longer} work when the mental rotation tasks are made closer to adults'.
\end{itemize}

\section{Abstract}

There is strong and diverse evidence for mental rotation (MR) abilities in adults. However, current evidence for MR in children rests on just a few behavioral paradigms adapted from the adult literature. Here, we leverage recent computational models of the development of children’s object recognition abilities to re-assess the evidence for MR in children. The computational models simulate infants’ acquisition of object representations during embodied interactions with objects. We consider two different object recognition strategies, different from MRs, and assess their ability to replicate results from three classical MR tasks assigned to children between the ages of 6 months and 5 years. Our results show that MR may play no role in producing the results obtained from children younger than 5 years. \review{In fact, we find that a simple recognition strategy that reflects a pixel-wise comparison of stimuli is sufficient to model children's behavior in the most used MR task.} Thus, our study reopens the debate on how and when children develop genuine MR abilities. 

\section{Introduction}

Mental rotation (MR) refers to the mental transformation of one object view into another view of the object \citep{shepard1971mental}. 
Performing MR is \review{thought} to be essential to recognize \review{a complex shape} from different viewpoints or to predict its appearance after a manual rotation. Thus, MR plays a key role in object perception and spatial reasoning. There is strong and diverse evidence for MR abilities in adults. In their seminal study \cite{shepard1971mental}, they presented adult participant \review{with pairs of objects composed of 1- one view of a complex 3D object and 2- either the same object rotated in-depth or a mirror-reversed version.} Importantly, only the 3D arrangement of basic features differs between the original and mirror versions of the object, limiting the cues available for recognition. The results showed that the recognition time was linearly dependent on the angle of rotation. After introspection, participants reported that they had to imagine an object as rotated in the same orientation as the original object to compare them. In general, these are strong arguments that support \review{the interpretation that MR involves a kind of mental simulation of physical rotations}. 

\review{Additional experiments have explicitly investigated the cognitive processes underlying MRs. These studies have shown that MR operates along the shortest angular path between two views \cite{searle2017mental}, and that this holds across different tasks and stimuli \cite{searle2017mental}. Together with the well-established positive correlation between angular disparity and reaction time, this suggests that MR involves a holistic transformation through the shortest rotational trajectory separating two viewpoints. Further research has proposed that MR reflects a covert simulation of motor actions. In a seminal study, \citet{wexler1998motor} introduced a dual-task paradigm in which participants were required to perform an MR recognition task while simultaneously rotating a joystick at a fixed speed, either clockwise or counterclockwise. The results show that incongruence between the direction of manual rotation and the direction of MR impairs both recognition accuracy and response time, whereas congruence facilitates performance. Furthermore, several studies report similar response patterns between mental and physical object rotations \cite{gardony2014does, wohlschlager1998mental}, and observed that increasing the angular disparity between two views is associated with increased activations in the parietal and premotor brain regions \cite{milivojevic2009functional, gogos2010greater}. Thus, in this paper, we assume that genuine MR involves a form of a stepwise mental transformation along a rotational path connecting two viewpoints.}

Understanding the \review{developmental origins} of MR requires an assessment of the use of MR in young children. The current evidence for MR in infants and toddlers rests mainly on two behavioral paradigms based on looking time, adapted from adult protocols. These tasks are made much easier by simplifying objects and providing additional cues to help children. For instance, these include a broader range of orientations of an object \cite{moore2008mental} or indirect information about the rotation applied between habituation and test images \cite{frick2014mental}. In addition, evidence of object recognition mostly relies on differences in looking time between an original object and its mirror version, without evidence for a linear relation between looking time and the angle of rotation. This is crucial as object recognition alone may also be the outcome of mental processes different from genuine MR \cite{tarr1989mental,de2000rotating,hayward2006dissociating,cheung2009dissociating}. For instance, children may simply compare object views based on low-level features like object apparent size or features that remain constant over different orientations. \review{To test MR abilities of children from 1 to 3 years old with similarly complex objects, a recent staircase procedure asks children to solve a rotation task with eye movements \cite{beckner2023novel}. However, there is no evidence of a linear relation between reaction time and rotation angle.} Overall, it remains unclear whether children rely on \review{genuine} MR to solve these tasks.

In this work, we use computational models to explore the role of simple object recognition strategies in solving standard MR tasks for children. We simulate three classical MR tasks based on 1- children's habituation; 2- adults' recognition and 3- children's violation-of-expectation. For these tasks, we model the recognition process with a \review{recent} bio-inspired machine learning model that learns visual representations \cite{aubret2024self}. It is trained by simulating child locomotion training, which is known to improve infants' abilities to solve these tasks \cite{frick2013development,schwarzer2013crawling,schwarzer2013crawling2,schwarzer2022locomotion,mohring2013touching,frick2014mental}. \review{After training, the vision model can simulate two recognition strategies used by humans}. 
The first strategy \cite{booth1998view,milivojevic2012object,tarr2017concurrent} simulates view-invariant \review{exemplar-based} object recognition by comparing \review{the similarity of} representations of different views without further mental processing. The second recognition process predicts the rotation transformation between two viewpoints of an object. \review{Importantly, we prevent the model from using implicit forms of MRs during this prediction}. %
Overall, this allows us to clearly identify which tasks need or do not need MR. To the best of our knowledge, this is the first computational model of object recognition used to \review{evaluate the necessity of genuine MR} in MR tasks for children.

\subsection{MR in preschoolers}

Starting at an age of 4--5 years, children can be evaluated based on their response times and provide an explanation of their strategy \cite{estes1998young}. Based on this paradigm, current studies indicate that 4--5 years is the onset of the development of MR abilities, with large individual differences. This is confirmed by their oral explanations of their reasoning \cite{estes1998young}, their reaction time to tasks similar to those of adults \cite{marmor1975development,marmor1977mental}, or when a child must rotate an object to fit into, e.g., an aperture \cite{noda2010manipulative,frick2013using,frick2013development}. Regarding manipulation tasks, another study found that 3-year-olds can perform MRs \cite{kruger2014analogue}; however, previous work highlighted that the use of MR may not be necessary to explain this result \cite{frick2014development}. More recently, other experiments \review{produced} similar findings with a simpler task, for which children had to decide whether an object matches a cutout, without explicit manipulations \cite{pedrett2023age}. In this work, we focus on assessing MR tasks for children younger than four years old.

\subsection{Assessing MR in young children with a habituation paradigm}

 A first approach to assess MR is to habituate infants to an object that rotates back and forth across a range of 240 degrees \cite{moore2008mental, schwarzer2013crawling, schwarzer2022locomotion,christodoulou2016seeing,constantinescu2018early,erdmann2018infants} . During testing, one measures the looking time when observing the same object, or its mirror version, rotating across the remaining $120^\circ$ angle. A difference in looking time is interpreted as the infant being able to recognize the original object by using MR. \review{Using this technique, evidence for MR} could be found in infants as young as 4 months \cite{slone2018object} or 3 months \cite{moore2011mental}.  These habituation tasks are much simpler than those of adults because the orientation angles of the habituation and test videos almost overlap. Rather than performing MR, \review{infants may attempt to compare their imperfect memory of object views at habituation orientations to views at similar test orientations}. Thus, instead of showing \review{continuous} videos, several studies habituated infants with \review{only 7 discrete} orientations of a ``1'' rotated \review{in-plane} by $45^\circ$ intervals \cite{quinn2014sex,quinn2008sex,kaaz2020infants}. Infants' looking time was higher on the eighth orientation for a symmetrical ``1'' \review{compared to} the original ``1''. \review{Note that the use of} in-plane rotations \review{provides a simpler setting, because these include no self-occlusions, e.g., a cup handle disappearing as the cup is rotated in depth}. To the best of our knowledge, \cite{gerhard2018impact} is the only habituation study to use videos \review{of in-depth rotations with a large} orientation gap of $54^\circ$ between habituation and test videos. They habituated infants on videos rotating in depth in a range of $180^\circ$; thereafter, they recorded the looking time of infants when rotating the same/mirrored object, by a previously unseen $90^\circ$ rotation. Looking time reflected that some infants differentiate the object from its mirror version. We analyze the role of MR in these studies in Section~\ref{sec:expe1}.

\subsection{Assessing MR in young children with a violation-of-expectation paradigm}

A second line of work studied the ability of young children to perform MR with a violation-of-expectation paradigm. In \cite{rochat1996tracking} and \cite{hespos1997dynamic}, a human was holding an object and performing an in-plane rotation while moving the object behind an occluder. When reappearing, the looking time of 6-month-old infants was higher when the reappearing object was improbable (symmetrical object), compared to when it was probable (same object). This suggests that infants were able to mentally rotate the object behind the occluder. In these experiments, the experimenter showed the beginning of the rotational movement of the object. A later study \cite{frick2013mental} investigated whether that movement is important for initiating a MR and found that infants who were given the opportunity to manipulate the objects prior to the experiment, looked longer at the improbable outcome than the probable one. This setting was adapted for in-depth rotation \cite{frick2014mental}. In this case, a common object was displayed on a turn table, was covered with a lid by an experimenter and revealed after a rotation around the yaw axis. In this case, the improbable outcome was an unexpected orientation of the object. Results showed that both 14-month-old who could manipulate the turntable prior to the experiment and 16-month-old without prior manipulation looked longer at the improbable outcome. We analyze the role of MR in these studies in Section~\ref{sec:expe2}.

\section{Experiment 1} \label{sec:expe1}

In this section, we aim to investigate whether the ability of young children to recognize an object, after habituation to videos showing its rotation, requires the use of MRs. We simulate the habituation paradigm used to evaluate in-depth MRs in previous studies \cite{moore2008mental,gerhard2018impact} and evaluate the ability of a computational model to solve the task without performing MR. Our hypothesis is that, for task settings used in the literature, MR is not necessary to recognize the object.

\subsection{Method}

\begin{figure}
    \centering
    \includegraphics[width=1\linewidth]{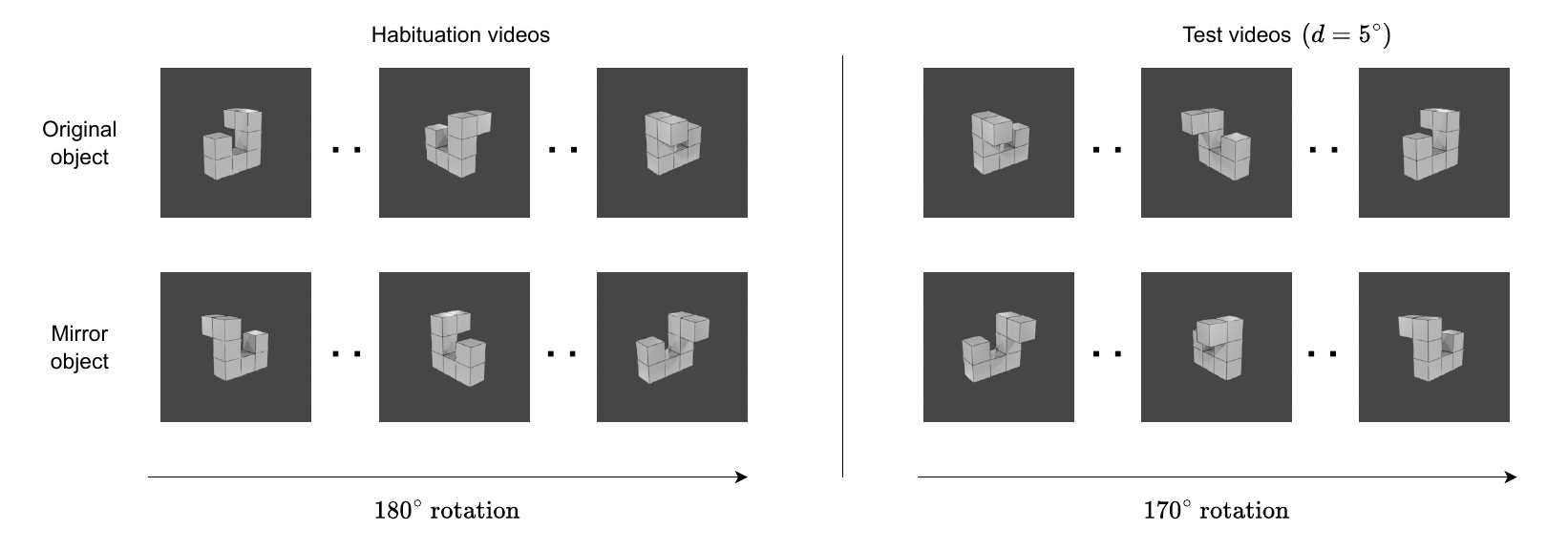}
    \caption{Examples of habituation and test videos for one original and mirror object. Here, we set $d=5$.}
    \label{fig:expe1}
\end{figure}

\paragraph{Stimuli} The habituation stimuli are Shepard Metzler shapes rendered with Blender in front of a dark grey background. We built 5 objects (``original'') along with their mirror version. We construct ``habituation'' videos by rotating the camera around an object from $0$ to $180^\circ$ and rendering the object every $5^\circ$. We similarly render the test images from $(180 + d)^\circ$ to $(360 - d)^\circ$ where $d \in [0, 80]$ defines the rotation gap between habituation and test images. $d$ controls the difficulty of recognition of the object. \figureautorefname~\ref{fig:expe1} illustrates the task. When $d=5$, there is almost an overlap between habituation and test videos, making the recognition task relatively easy; when $d=50$, \review{habituation and test stimuli are visually more distinct,} and this matches more closely the experiments of \cite{gerhard2018impact}. In practice, we replicate the habituation and test videos for 8 different ranges of original orientations shifted by $45^\circ$, with both the original and the mirror objects. This leads to a total of $8 \times 5 \times 2 = 80$ pairs of habituation/test videos for a given rotation gap $d$.

\paragraph{Computational model} 

\review{We aim to simulate one of the main alternatives to MR for recognizing objects, namely the use of view-invariant object representations \cite{tarr2017concurrent,booth1998view}.
Thus, we use AA-SimCLR \cite{aubret2024self}, a bio-inspired model, based on deep neural networks, that builds on SimCLR \cite{chen2020simple}. SimCLR learns, without supervision, image representations that are invariant to geometric and color transformations of images. This category of unsupervised models is called self-supervised \cite{chen2020simple}. AA-SimCLR further implements two principles of biological learning. First, AA-SimCLR implements the biological principle of temporal slowness \cite{li2008unsupervised,franzius2011invariant,foldiak1991learning,aubrettime}, so the model learns similar representations for close-in-time visual inputs. When trained with simulted children's natural interactions with objects, this leads to visual representations that are invariant to the viewpoint of an object \cite{aubrettime,schaumloffel2023caregiver}. This enables simulating object recognition with a strategy that relies on constructing view-invariant object representations (Experiments 1 and 2). Second, AA-SimCLR implements sensorimotor learning \cite{o2001sensorimotor} by learning representations of the action applied between two images. More specifically, AA-SSL aligns the quaternion of the camera rotation with a projection of the pair of images before and after the rotation. This projection is composed of 3 neural network layers. When the actions are egocentric rotations around an object, AA-SimCLR learns representations of a rotation transformation. That way, the model can perform rotation recognition to model children's looking times, which will be useful in Experiment 3. To emulate the visual experience of a child who crawls/walks around objects, which is thought to be important for solving MR tasks \cite{frick2013development,schwarzer2013crawling,schwarzer2013crawling2,schwarzer2022locomotion,mohring2013touching,frick2014mental}, we pre-train AA-SimCLR on MVImgNet \cite{yu2023mvimgnet}, a large-scale dataset of 219,188 real-world videos showing rotations around an object. The model is pre-trained in an unsupervised fashion for 30 epochs with a ResNet18, the AdamW optimizer, a batch size of 256 and a learning rate of $0.001$. Other hyperparameters are identical to those in the original paper.}

\paragraph{Model of habituation task} During pre-training, the model learns to extract similar abstract representations for different views of the same object. To take advantage of this property in our recognition task, we model object recognition as a simple comparison between habituation and test image representations. \review{Following standard practices \cite{chen2020simple,aubret2024self}, we define visual representations as neuron activations immediately after the average pooling layer of ResNet18.} This allows us to assess whether learning of view-invariant visual representations is sufficient to solve the given recognition task. To simulate a form of novelty preference, we follow three steps. \review{First, we compute the pairwise cosine similarities $s_{i,j}$ between the representations of the habituation images of an object $x_i$ and the representations of test orientations of the same object $x_j$. Here, $i \in I,\, \text{where } I = [s,s+180]$ refers to the habituation orientations, with $s$ being the original initial orientation. In addition, $j \in J,\, \text{where } J = [s + d, s + 180 - d]$ refers to the test orientations given a difficulty $d$. We similarly compute the set of similarities $s'_{i,j}$ between habituation images and a test image of the mirror object $x'_j$. Second, we compute two different familiarity scores based on two opposite variants of the exemplar theory of categorization \cite{nosofsky1991tests,lamberts1995categorization,nosofsky2002exemplar,medin1978context}. Assuming that children approximately compute the global similarity between habituation and test images (``Average'') \cite{medin1978context,nosofsky1986attention}, we compute the familiarity scores for test images of the original objects \[F(x_j)= \frac{1}{|I|\,|J|} \sum_{i \in I} \sum_{j\in J}s_{i,j}\]  and mirror objects \[F(x'_j)= \frac{1}{|I|\,|J|} \sum_{i \in I} \sum_{j\in J} s'_{i,j}. \] Assuming instead that children approximately employ a winner-take-all strategy for memory retrieval of the closest exemplar image (``Maximum'') \cite{lamberts1995categorization}, we compute the familiarity scores for test images of the original objects $F(x_j)= \max_{i \in I,\, j \in J} (s_{i,j})$  and mirror objects $F(x'_j)= \max_{i \in I,\, j \in J} s'_{i,j}$.} Third, we define a novelty preference score of an object as the familiarity score being higher for the original object compared to its mirror version, \review{\textit{i.e.} $F(x_j) > F(x'_j)$}. We average the results over objects and the different orientations described in the Stimuli section.

\subsection{Results}

\begin{figure}
    \centering
    \includegraphics[width=1\linewidth]{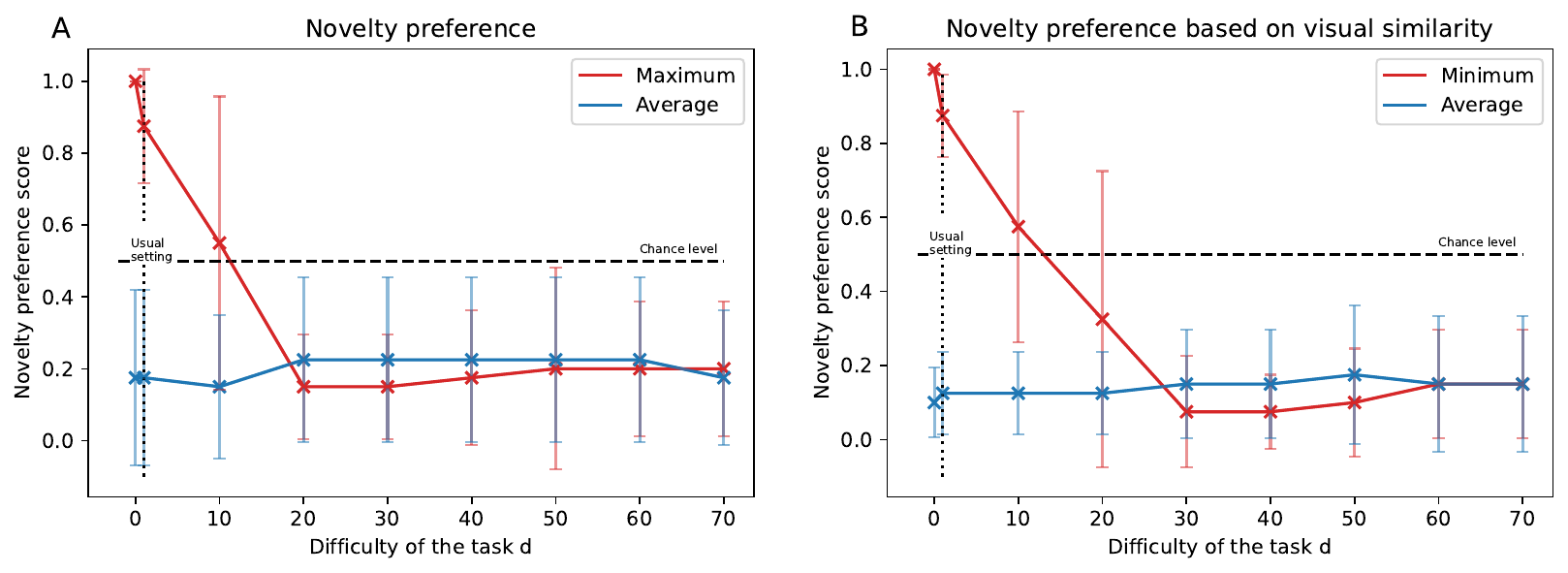}
    \caption{A) Novelty preference score of the computational model. We compute the mean and standard deviation of the score across objects. ``Usual setting'' refers to the value used in previous publications \cite{moore2008mental,schwarzer2022locomotion}. B) Portion of views that are closer to their rotated version than to the rotated mirror, based on Euclidian distance between pixels.}
    \label{fig:familiaritypref}
\end{figure}

In \figureautorefname~\ref{fig:familiaritypref}A), we show the novelty preference score of our computational model.  We first see that the model presents a high novelty preference score, under the ``maximum'' comparison assumption, when the task difficulty is comparable to, and even higher than, that of previous studies \review{claiming evidence for} MR in children ($d \in [0, 20]$). This suggests that success in solving this task may reflect mental processes different from those involved in MRs, \review{such as exemplar-based object recognition}. \review{Our model solves the task without MR but instead by} looking for the two closest representations of views in the habituation and test videos. \review{However, we do not observe this effect for the ``average'' comparison assumption. This indicates that the exact exemplar strategy used matters. While we are unaware of studies that provide evidence favoring one strategy over the other in young children, our result shows that the ``Maximum'' strategy better fits child data than the ``Average'' strategy.} Importantly, we observe that as difficulty increases ($d > 20$), the novelty preference score quickly decreases. Thus, MR may be necessary to solve more complex versions of the task.

For harder tasks ($d > 20$), the present model finds that the habituation stimuli are more similar to a mirror rotated objects than the original rotated object. Similar results were observed in a previous study with children \cite{gerhard2018impact}. While children looked longer at mirror objects when $d=1$, children looked longer at the familiar object for $d=54$. The authors interpreted that as an internal switch from novelty preference to familiarity preference. In contrast, our results suggest that the features extracted from the test images of the original object may be more novel than those of the novel object.

\review{Next, we hypothesized that a novelty preference over perceptual visual features could induce this effect. For example, the visual size of an object at $0^\circ$ may closely resemble that of its mirror at $180^\circ$. We test this hypothesis in \figureautorefname~\ref{fig:familiaritypref}B), where we compute the novelty preference score using the Euclidian distance between the images' pixels as the similarity metric. Despite the basic visual comparison, the preference score based on visual similarity presents a similar trend as the novelty preference score based on the model of view-invariance. Similarly to children, the computational model was trained on common objects; this suggests that the particular design of Shepard-Metzler shapes can also be challenging for a model of view-invariant object learning. Overall, our study suggests that child looking time for harder tasks ($d > 20$) could result from a cognitive comparison that reflects the perceptual novelty of a view.}

\section{Experiment 2}

The previous experiment showed that simply comparing views of an object is sufficient to model children's behavior during an MR task. Here, we aim to assess whether the object recognition mechanism presented in Experiment 1 still works for the tasks used to evaluate MR in adults \cite{shepard1971mental}. We consider a variant of Experiment 1 that replicates MR tasks with adults. Specifically, we replace habituation and test videos by single views of Shepard-Metzler objects and keep the rest unchanged. %

\subsection{Results}

\begin{figure}
    \centering
    \includegraphics[width=0.8\linewidth]{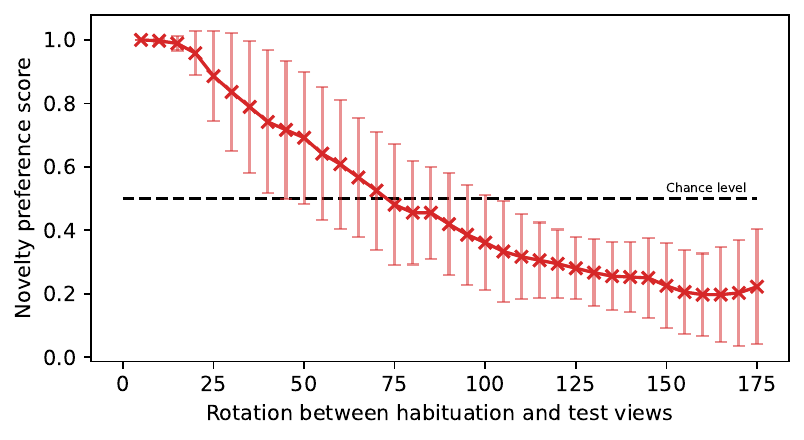}
    \caption{Novelty preference score of the computational model on an adults-based variant of Experiment 1. The rotation angle refers to the angle difference between the first and rotated view of the object. }
    \label{fig:resexpe3}
\end{figure}

\figureautorefname~\ref{fig:resexpe3} show the results of the computational model. We observe that simply comparing the visual representation allows the model to solve the task when the object is slightly rotated ($r < 50^\circ$. We also find a similar effect as in Experiment 1, for which the model shows a higher novelty preference score for the familiar object for large rotations. Interestingly, we also find a strong linear correlation between the rotation angle and the novelty preference score (Pearson coefficient $-0.96$, $p=10^{-20}$), which reflects the increasing difficulty of the task. We conclude that the model cannot reliably solve this task. More sophisticated strategies, such as genuine MR, may be necessary \review{to explain human behavior in this task}.

\section{Experiment 3}  \label{sec:expe2}

In this section, we explore whether the ability to recognize an object, after being hidden behind \review{an occluder} and rotated, requires the use of MR. We simulate the violation-of-expectation used to evaluate in-depth MR in a previous study \cite{frick2014mental} and evaluate the ability of a computational model to solve the task without performing MR. We hypothesize that the experiment may violate an expectation about the rotation transformation of the object \review{(``a rotation of $80^\circ$ was applied between the two views'') rather than the outcome of this transformation (``the view should resemble that after a rotation'')}.

\subsection{Method}

\paragraph{Stimuli} The original experiment used common toys \cite{frick2014mental}. In the present study, we use the OmniObject3D dataset, a collection of photo-realistic 3D models of common objects rendered from 24 random viewpoints \cite{wu2023omniobject3d}. In practice, we leverage a preprocessed subset containing 1434 objects, used for evaluating models \cite{aubret2024perceive}. We record three views per object, following \cite{frick2014mental}. The first view presents the object facing the camera. To simulate a physically coherent rotation $r$ of the hidden object, the second image shows the closest view to a \review{clockwise} rotation of the object by $r \in [5, 175]^\circ$ (``probable view''). To simulate an improbable outcome of a hidden rotation of the object, The third view shows the closest view to a \review{counterclockwise} rotation of the object by $-r$ (``improbable view''). \review{Thus, the probable and improbable views result from the same rotation magnitude but applied in opposite directions.}

\paragraph{Computational model}
\review{Our model of the task requires a computational model that can infer the rotation transformation applied between two viewpoints. Thus, we use the same computational model, AA-SimCLR, as for Experiments 1 and 2. This task considers common objects, which allows us to use a simpler training dataset to simulate child exposure to different views of objects. We train the model on a train split of OmniObject3D \cite{wu2023omniobject3d} that was used in \cite{aubret2024perceive}. This dataset does not present backgrounds, contains 6,000 full rotations of objects around the yaw axis. For evaluation (see below), we use the test split of this dataset, allowing us to bypass the well-known machine learning issues that arise when viewing very different data at train and test time. Here, the sine and cosine of the rotation angle between two views \review{are used as the ``action'' input} of AA-SimCLR. We use the same hyperparameters as in Experiment 1.}

\begin{figure}
    \centering
    \includegraphics[width=0.8\linewidth]{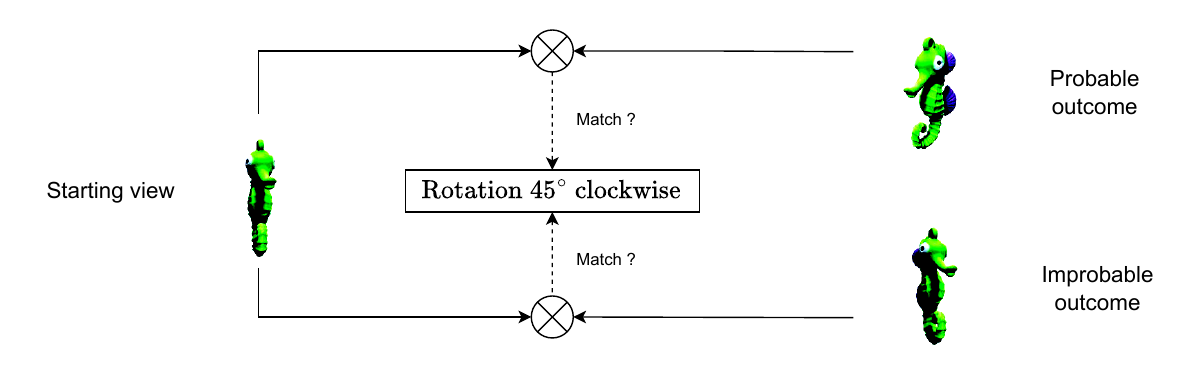}
    \caption{Illustration of how we compute the consistency scores. We quantify how much the rotation transformation predicted from two views of the object, before and after transformation, matches the true rotation transformation. In this example, the object was turned by $r=45\degree$.}
    \label{fig:voe}
\end{figure}

\paragraph{Model of Violation-of-expectation task} During pre-training, the model \review{was trained} to \review{predict the rotation transformation executed to transition from one view to another}. Thus, the model forms expectations about the rotation transformation itself. \review{Since the model uses only 3 sequential processing steps (neural network layers) to infer the rotation transformation, it can not implicitly perform a MR: MR requires one to process step-by-step the path between views.} We assume that the model and participants in previous studies access the true rotation transformation: the original studies show several cues indicating the true rotation transformation between two views (turning table \cite{frick2014mental}, object motion \cite{rochat1996tracking} etc.). To model a violation of expectation of rotation transformation, we define the consistency score as the cosine similarity between the true rotation transformation and the rotation transformation extracted by our model from two object views. We compute this consistency score for the original and probable view, as well as for the original and improbable view. \figureautorefname~\ref{fig:voe} illustrates the computation. Finally, we compute the rotation recognition accuracy as the proportion of objects for which the consistency score of the probable view is higher than the improbable view.

\subsection{Results}

\begin{figure}
    \centering
    \includegraphics[width=1\linewidth]{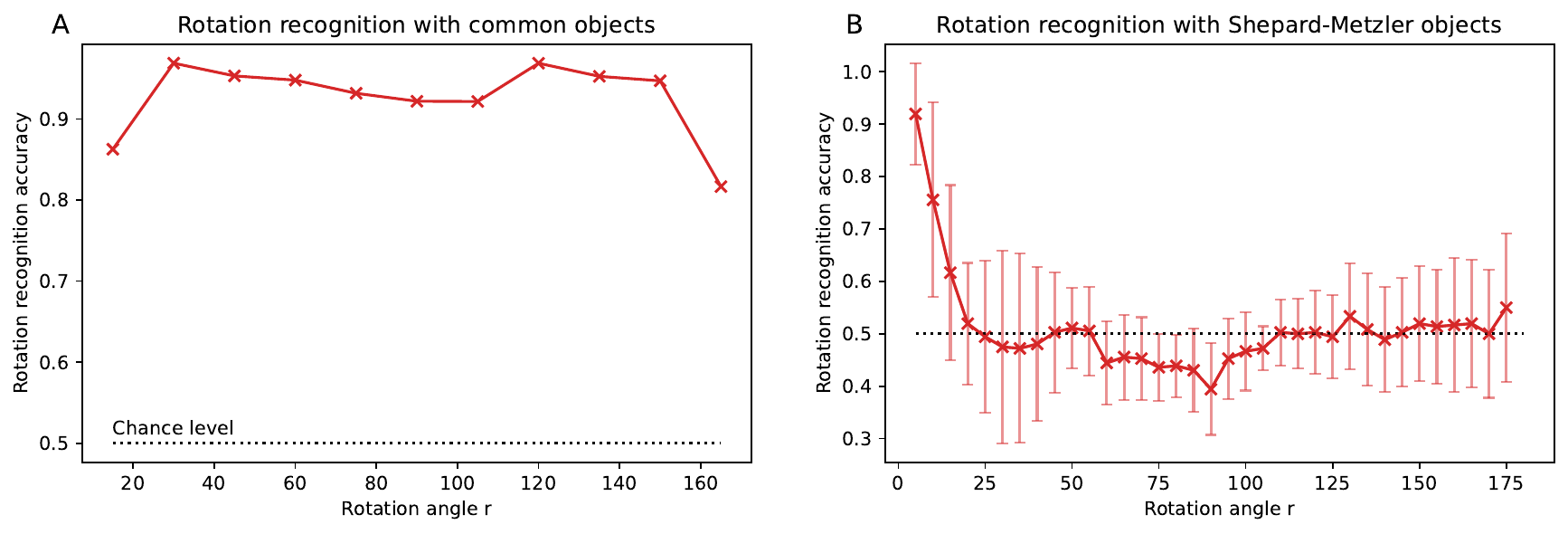}
    \caption{Rotation recognition accuracy with A) common objects and B) Shepard-Metlzer objects. Error bars in B) indicate the standard deviation across objects.}
    \label{fig:resexpe2}
\end{figure}

In \figureautorefname~\ref{fig:resexpe2}A), we show the rotation recognition accuracy of our computational model. We clearly observe that the model recognizes the true rotation transformation with high accuracy (close to 1), regardless of the rotation angle of the simulated hidden rotation. Interestingly, the model performs worse for angles close to $0^\circ$ and $180^\circ$. This is presumably because the probable and improbable views are more visually similar to each other in these situations. Overall, we conclude that a simple mechanism based on the recognition of a rotation transformation can also explain the performance of children in previous studies \cite{frick2014mental}.

Next, we increase the complexity of the task to \review{test} the limits of this recognition strategy. We replicate the experiment but replace the common objects by the Shepard-Metzler objects used in Experiments 1 and 2. Then, we define the improbable outcome as a mirror object. In \figureautorefname~\ref{fig:resexpe2}B), we see that the model does not recognize the rotation transformation for angles larger than $10^\circ$. \review{We suspect that the curated complexity of Shepard-Metzler objects makes this recognition strategy obsolete.} This suggests that more sophisticated recognition strategies, like MR, may be crucial to solving a violation-of-expectation task with Shepard-Metzler objects.

\section{Discussion}

The objective of this study was to re-assess the evidence for MR abilities in children. We simulated three classical tasks used to study MR in young children and adults, including a habituation task for children, a recognition task for adults and a violation-of-expectation task for children. We modeled two object recognition processes with a computational model that was pre-trained in a bio-inspired fashion. Our results show that a simple comparison of visual representations explains the results of children in a classical habituation task. \review{We find that these results reflect the ability of the model to capture the visual similarity of views}. In addition, extracting the rotation transformation applied between two views of an object explains the results of children in a violation-of-expectations task. Although this does not rule out the use of MR by children, this result advocates experiments that control for other object recognition mechanisms.

\review{In the second experiment, we found that our model was unable to solve MR tasks typically used with adults, likely reflecting the essential role of MR in successfully performing these tasks. Our experiments introduced intermediate conditions, between the original tasks designed for young children and those for adults, that the non-MR model also failed to solve. For habituation tasks, our results suggest that increasing the angular disparity to at least $20^\circ$ may require MR for successful object recognition. In violation-of-expectation tasks, substituting familiar objects with classical Shepard-Metzler shapes significantly impaired model performance, particularly when the rotation angle exceeded $20^\circ$. Based on these findings, we recommend that minimally constrained conditions to test MR in young children should include Shepard-Metzler shapes and angular disparities greater than $20^\circ$.
}

 To the best of our knowledge, this is the first computational model of object recognition used to \review{rigorously test the necessity of a MR mechanism in MR tasks designed} for children. However, the range of tasks used to study MR in children is broad, and we focused on a representative subset of these experiments. First, we simulated tasks related to in-depth MRs around the yaw axis, mostly because the datasets used for pretraining our computational models mainly display in-depth rotations of objects. Other studies with children applied in-plane rotations \cite{quinn2008sex,quinn2014sex} or in-depth rotations around the pitch axis \cite{gerhard2018impact}. How this choice impacts the results of our computational models (and of children) is currently unknown. However, we suspect that the recognition of an object after an in-plane rotation may also be easier than after an in-depth rotation, \review{making MRs partially unnecessary \cite{muto2021evidence,kung2010model}.} \review{In-depth rotations pose a harder challenge, because they can produce object parts becoming hidden due to self-occlusions}. 
 Second, we simulated experiments that used Shepard-Metzler objects or common objects, but other experiments used other types of stimuli, such as numbers \cite{quinn2008sex,quinn2014sex}. One cannot rule out that children had prior exposure to these numbers, which may help them to recognize them with non-MR strategies when rotated.

 This work paves the way towards the use of computational models for understanding the advantages and developments of different strategies of object recognition. For now, the pre-training phase of our computational model simulates the experience and training of children before the experiments. However, the computational model trains on more interactions with objects than what a young child may experience. Assessing the role of different recognition mechanisms according to different levels of prior experience may provide new insights into their role in object recognition tasks. In addition, the computational model currently lacks the ability to explicitly simulate mental rotations, which may promote even better recognition scores than the investigated recognition mechanisms. So far, there are several candidate learning models for \review{explicitly} simulating MRs \cite{zhou2024dino,hansentd}, but they have never been trained on real-world object images.

 Taken together, our results support that young children, unlike adults, may employ non-MR recognition mechanisms to solve \review{the MR tasks administered to them}. This calls for new experimental designs that can detect MR at different ages and a review of the different factors linked to the development of MR, such as the origin of sex differences in MR \cite{moore2020development,linn1985emergence,geiser2008note,voyer1995magnitude}. Future computational models will need to model MR and more accurately simulate the exact level of prior knowledge in young children.

\section*{Acknowledgement} This work was funded by the Deutsche Forschungsgemeinschaft via the project ``Abstract REpresentations in Neural Architectures'', as well as the projects ``The Adaptive Mind'' (Project No. 533717223) and ``The Third Wave of Artificial Intelligence'' funded by the Excellence Program of the Hessian Ministry of Higher Education, Science, Research and Art (HMWK). JT was supported by the Johanna Quandt foundation. The authors gratefully acknowledge the computing time provided to them at the NHR Center NHR@SW at Goethe University Frankfurt (project autolearn). This is funded by the Federal Ministry of Education and Research, and the state governments participating on the basis of the resolutions of the GWK for national high performance computing at universities (www.nhr-verein.de/unsere-partner). This work was also supported in part by computational resources provided by the MaSC high-performance computing cluster at Philipps-Universität Marburg.

\bibliographystyle{apalike}
\bibliography{main}

\end{document}